\documentclass[conference, letterpaper, 11pt]{IEEEtran}
\IEEEoverridecommandlockouts
\newcommand{\CLASSINPUTtoptextmargin}{0.75in}
\newcommand{\CLASSINPUTbottomtextmargin}{0.78in}
\newcommand{\CLASSINPUTinnersidemargin}{0.625in}
\newcommand{\CLASSINPUToutersidemargin}{0.625in}
\usepackage[utf8]{inputenc}
\usepackage{cite}
\usepackage{amsmath,amssymb,amsfonts}
\usepackage{graphicx}
\usepackage{textcomp}
\usepackage{xcolor}
\usepackage{url}
\usepackage{colortbl}
\usepackage{tabularx}
\usepackage{array}

\begin{document}
\setlength{\columnsep}{0.25in}
\addtolength{\topmargin}{0.04in}
\title{Enhancing Edge AI System Security with a Stacked Hardware Security Controller for Robust and Energy-Efficient Anomaly Detection}

\author{\IEEEauthorblockN{Carl Benda}
\IEEEauthorblockA{North Carolina State University\\
cmbenda@ncsu.edu}
\and
\IEEEauthorblockN{Eren Kurshan}
\IEEEauthorblockA{Princeton University\\
ekurshan@princeton.edu}
\and
\IEEEauthorblockN{Paul D. Franzon}
\IEEEauthorblockA{North Carolina State University\\
paulf@ncsu.edu}
}

\maketitle

\nocite{*}

\begin{abstract}
In recent years, the growing adoption of Edge AI systems has been accompanied by a surge in security threats. Edge AI deployments face a rapidly evolving spectrum of security threats, from side-channel information leakage to adversarial model manipulation and physical device tampering. 
Conventional security mechanisms such as malware scanners and resource-intensive firewall solutions are unsuitable due to edge AI's typically limited computational resources, strict energy budgets and thermal constraints especially in battery-powered and field-deployed environments.

This paper proposes a stacked, hardware-based security mechanism that enables low-overhead, continuous detection of anomalous execution patterns. Anomaly detection utilizing hardware performance counters and physical sensors outperforms software-based alternatives by providing key capabilities such as tamper-resistant root-of-trust isolation, negligible runtime overhead, and resilience against compromised operating systems. Experimental analysis shows that adversarial or malicious applications produce observable deviations in sensor and thermal readings on Edge AI systems and hardware-based anomaly detection provides effective and low-overhead anomaly detection.
\end{abstract}


\section{Introduction}
Edge AI systems have witnessed substantial growth recently \cite{r20,r21},
characterized by significant advancements in computational capabilities and an increasing number of deployments across diverse application domains \cite{r1,r5}. Their current deployments span numerous domains including autonomous vehicles, smart cameras, wearable medical devices, robotic platforms and smart city infrastructures. These cyber-physical systems utilize a wide range of hardware devices from smaller battery powered sensor devices to higher power processors and graphics processing units.

The rapid growth in the number of deployments has also fueled a large number of novel cyberattacks, ranging from physical device access-based attacks and side-channel attacks \cite{r26}
to snooping attacks in multi-access edge devices \cite{r9,r11,r13}. Despite this increase in the number and diversity of cyberattacks, limitations in power, thermal, and computational budgets of edge AI devices challenge the feasibility of conventional software-based security solutions, posing a two-pronged challenge.

In this paper, we propose a stacked hardware security unit to continuously monitor and detect malicious and anomalous behavior with minimal overhead using hardware sensor infrastructure capabilities. Stacked hardware security unit provides fine-grain local access to the underlying hardware macros and sensors by sitting on top of the Edge AI device layer. Furthermore, unlike software-dependent methods, hardware-backed anomaly detection that leverages performance counters and physical sensors (i) achieves superior efficacy by establishing tamper-resistant hardware isolation, (ii) introduces minimal runtime overhead, and (iii) maintains resilience against compromised host edge AI operating systems.
Beyond conventional micro-architectural monitoring, the proposed AI Guard-Layer architecture integrates advanced semantic activity tracking and multi-layered verification mechanisms, establishing a comprehensive defense framework for edge AI deployments \cite{r19}.

Our experimental analysis shows that malicious applications generate distinct thermal profiles on edge AI architectures. Using underlying hardware and thermal sensors, the stacked security architecture discriminates between normal behavior and anomalous miner application behavior. The paper is organized as follows: Section II discusses the background and the growing cybersecurity challenges; Section III proposes a stacked hardware security controller approach; Section IV presents and discusses experimental analysis and results; and finally Section V provides concluding remarks.

\section{Edge AI Security}

\subsection{Background and Security Challenges}

With the rapid growth in edge AI deployments \cite{r5,r8}, security challenges have proliferated \cite{r1,r3,r22}.
A wide range of attack types have been reported from adversarial \cite{r23}
and trojan attacks \cite{r4}, malware attacks \cite{r9} to data poisoning \cite{r7,r10} and DDoS \cite{r6}. In addition, a large number of application specific cyberattacks have been targeting smart vehicles, drone systems \cite{r11}, robotic systems, wearable medical devices, intelligent IoT devices and other edge AI systems.

Despite the exacerbating cybersecurity challenges, edge AI devices have serious resource limitations \cite{r14,r15} in contrast to the standard cloud-based AI systems: \\
\noindent \textit{(i) Energy and Battery Limitations:} Edge AI devices frequently operate on battery power and have severe energy constraints, which limit their ability to run energy hungry security software and applications. \\
\noindent \textit{(ii) Communication Bandwidth:} Similarly, communication bandwidth is constrained compared to the cloud-based AI systems, which limits the network-level security solutions.\\
\noindent \textit{(iii) Thermal Packaging Constraints:} As edge AI devices are often embedded or wearable systems, they are often physically constrained to fully sealed, or passively cooled enclosures designed to survive harsh ambient conditions. \\
\noindent \textit{(iv) Computational Constraints:} Edge AI devices have limited computational power to continuously monitor malicious activity, which becomes a bottleneck for software-based techniques.\\
\noindent \textit{(v) Physical Device Security:} As edge AI devices are deployed in the field and often in adversarial environments, they are nearly impossible to fully physically secure, which makes it easier to install and run malicious code that conceals its activity patterns.

As a result, traditional software-based security solutions are not effective in edge AI systems. Their large overheads pose serious risks for full battery depletion and brownout, which causes degraded output quality due to resource constraints and trade offs. In addition, a large number of cyberattacks including DDoS, malware and trojan attacks throttle and strain the underlying hardware resources. This often reduces the bandwidth available for detection and remediation, but it provides a unique opportunity to detect anomalous activity patterns.

\section{Stacked Security Layer}

Active 2.5D interposer and 3D integrated packaging, commonly used in embedded systems and edge AI solutions, provide a unique opportunity to address these challenges \cite{r2,r16,r17}. Stacking provides local and direct access to the underlying edge AI hardware through TSV or micro-bump level connections. This high-bandwidth, highly-localized access provides a chance to closely monitor the edge AI system at a fine-granularity (at the hardware macro level) through both hardware activity counters and physical (thermal and electromagnetic) monitoring perspective.

\begin{figure}[!t]
\centering
\includegraphics[width=\columnwidth]{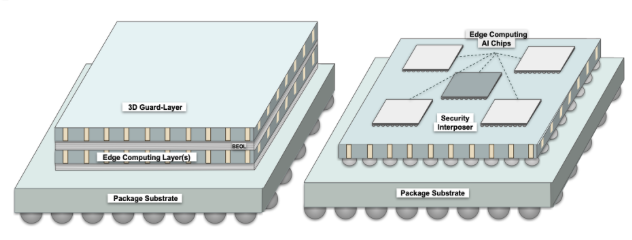}\\[2pt]
{\footnotesize\makebox[0.5\columnwidth]{(i)}\makebox[0.5\columnwidth]{(ii)}}
\caption{(i) 3D guard-layer configuration with one or more edge intelligence layers and stacked 3D guard-layer, (ii) 2.5D Silicon interposer configuration with multiple edge intelligence chips and a dedicated hardware security guard-layer interposer connected to all AI chips.}
\label{fig1}
\end{figure}

In this paper, we propose a stacked hardware security architecture (2.5D or 3D) vertically integrated with the primary edge AI chip. This stacked controller: (i) continuously monitors on-chip activity directly and indirectly through activity and hardware counters, thermal sensors, noise sensors and other sensors (ii) continuously learns from underlying hardware activity from the streaming data, (iii) detects anomalous activity by classifying normal and anomalous activity patterns locally with minimal performance overhead.

In addition, the stacked cybersecurity architecture provides additional capabilities such as (i) physical guarding against side-channel attacks through 3D stacking (ii) shadow processing of the streaming data for data poisoning prevention (iii) AI safety monitoring \cite{r12,r13} (iv) redundant processing capabilities for hardware resilience. Figure 1 shows the stacking configurations using 3D integration with the edge AI hardware security controller (i) and 2.5D interposer version (ii) to monitor multi-chip edge AI systems.

\section{Experimental Analysis and Results}

\subsection{Experimental Setup}

Hardware telemetry was collected on an NVIDIA L40S GPU (48 GB GDDR6, 350 W TDP) allocated via the NCSU HPC cluster (IBM Spectrum LSF scheduler, short\_gpu queue). L40S was selected for its use in a range of edge AI applications. All jobs were submitted with \textit{mode=exclusive\_process} GPU allocation to prevent co-tenant interference, an important control given that prior runs using shared scheduling mode produced contaminated idle data with 99\% spurious SM utilization from co-scheduled user jobs.

Data collection used a custom NVML-based collector (Python/pynvml) operating at 10 Hz, capturing 20 hardware telemetry metrics per polling cycle: GPU temperature, power draw, SM and memory utilization, SM and memory clock frequencies, PCIe RX/TX throughput, fan speed, ECC single-bit and double-bit volatile error counters, and supporting identifiers. All data was written directly to Parquet on cluster-attached storage without network dependency, ensuring no collection overhead on the GPU under test. The following workload conditions were collected on NVIDIA L40S GPUs of the same model and configuration:

\begin{table}[!t]
\caption{Descriptions for Idle, Training, Inference, Proxy Miner, DDoS and RowHammer workloads used in experimental analysis.}
\label{tab1}
\centering
\definecolor{tablehdr}{RGB}{46,52,64}
\definecolor{normalrow}{RGB}{200,255,200}
\definecolor{anomalyrow}{RGB}{255,200,200}
\renewcommand{\arraystretch}{1.25}
\setlength{\tabcolsep}{1pt}
\footnotesize
\begin{tabularx}{\columnwidth}{|>{\bfseries}p{1.6cm}|p{1.1cm}|p{1.5cm}|>{\raggedright\arraybackslash}X|}
\hline
\rowcolor{tablehdr}
{\color{white}\textbf{Workload}} & {\color{white}\textbf{Class}} & {\color{white}\textbf{--mode flag}} & {\color{white}\textbf{Description}} \\
\hline
\rowcolor{normalrow}
Idle & Normal & \texttt{idle\_0pct} &
  GPU reserved (\textit{excl.\ proc.}), no compute process.
  Minimum thermal and power floor. No \texttt{gpu\_stress.py} --- collector only. \\
\hline
\rowcolor{normalrow}
Training (100\%) & Normal & \texttt{-{}-mode train} &
  Sustained FP32 matrix-multiply ($8192{\times}8192$) via CuPy, 300\,s.
  Saturates all 18{,}176 CUDA cores. Representative of DNN training. \\
\hline
\rowcolor{normalrow}
Inference (75\%) & Normal & \texttt{-{}-mode infer} &
  Two-layer MLP forward pass ($2048{\times}2048$), 40\,ms inter-batch sleep.
  Lower, burstier SM activity than training. \\
\hline
\rowcolor{anomalyrow}
Proxy Miner & Anomaly & \texttt{-{}-mode train\,*} &
  Same CUDA matrix-multiply kernel as Training, divergent SM scheduling behavior;
  labels Parquet as \texttt{workload=miner}. 0\% SM util despite elevated power
  and pinned clocks.\textit{ * No \texttt{-{}-mode miner} in \texttt{gpu\_stress.py}.} \\
\hline
\rowcolor{anomalyrow}
DDoS/Exfil & Anomaly & \texttt{-{}-mode ddos} &
  Repeated 1\,GB GPU$\rightarrow$CPU DMA via pinned memory. No compute kernels.
  Simulates model-weight exfiltration via CUDA DMA engine. \\
\hline
\rowcolor{anomalyrow}
RowHammer & Anomaly & \texttt{-{}-mode rowhammer} &
  CUDA kernel hammers DRAM rows via volatile loads. 192\,MB buffer exceeds
  48\,MB L2, forcing row activations. Double-sided DRAM attack~\cite{r24,r25}. \\
\hline
\end{tabularx}
\end{table}

\textit{(i) Idle:} No user processes attached. GPU in driver-managed power state.
300-second collection, 59,660 metric-level data points (20 channels, 2,983 polling cycles).

\textit{(ii) Training:} Sustained FP32 matrix-multiply (8192$\times$8192) via CuPy, 100\% intensity with no sleep pacing. 300-second collection, 59,400 metric-level data points.

\textit{(iii) Inference:} Batch inference simulation: two-layer MLP forward pass (2048$\times$2048) with 40 ms inter-batch sleep gaps. Produces lower, burstier SM activity than training.

\textit{(iv) Proxy Miner:} Workload structured to emulate cryptomining behavior. The compute pattern differs from the training workload in its SM scheduling behavior, as described in Section~\ref{sec:results}. 300-second collection, 59,840 metric-level data points.

\textit{(v) DDoS:} Repeated 1 GB GPU$\rightarrow$CPU DMA transfers using pinned memory. No compute kernels. Simulates covert model-weight exfiltration via CUDA DMA engine.

\textit{(vi) RowHammer:} Custom CUDA kernel hammers aggressor DRAM rows using volatile loads. 192 MB buffer exceeds 48 MB L2 cache, forcing DRAM row activations. Simulates double-sided DRAM attack \cite{r24,r25}.

Data collection used 10 Hz polling. Temporal stability (T1) was verified via coefficient of variation (CoV = std/mean $\times$ 100\%) for temperature across each run. All conditions except RowHammer (CoV = 11.46\%) achieved CoV $<$ 9\%; note that for workloads with an initial thermal ramp (Training, RowHammer), CoV is computed over the full 300-second window. ECC volatile error counters were zero across all conditions. Hardware metrics were collected at user privilege via NVML without root access, consistent with the threat model in which a hardware security monitor operates within normal OS privilege boundaries.

\subsection{Experimental Results}
\label{sec:results}

We collected hardware telemetry traces on an NVIDIA L40S GPU \cite{r18} across six workload conditions: idle, sustained matrix-multiply training at 100\% intensity, inference, a proxy cryptominer workload, DDoS and RowHammer. All traces were collected at 10 Hz via NVML using exclusive GPU mode to eliminate co-tenant interference. Twenty hardware metrics were captured per polling cycle including temperature, power draw, SM and memory clock frequencies, SM utilization, PCIe throughput, and ECC error counters.

\textit{Idle Baseline:} At idle the L40S maintained a stable thermal floor of 29.8$^{\circ}$C (CoV = 2.4\%) at 44.6 W mean power, with SM clock cycling dynamically between 0 and 1755 MHz. This establishes the reference distribution against which anomalous workloads are detected.

\textit{Legitimate Training Workload:} Sustained matrix-multiply execution produced an immediate power step to 348.9 W and a characteristic thermal ramp from the idle floor to a steady-state equilibrium of 66.0$^{\circ}$C over approximately 120 seconds. $+36.2\,^{\circ}$C delta is shown above the idle baseline. SM utilization registered 99.7\% throughout, SM clock stabilized at 1478 MHz, and PCIe RX throughput increased to 524 KB/s. These signatures collectively characterize a legitimate high-intensity compute workload.

\textit{Proxy Miner (Malicious Workload):} The miner workload produced a qualitatively different profile: power stabilized at 87.2 W (CoV = 0.5\%), double the idle baseline but one-quarter of the training load, while SM utilization reported 0\% throughout the entire collection window. Despite zero reported SM utilization, the SM clock held at a mean of 2422 MHz (range 2370--2520 MHz) and memory clock locked at 9000 MHz, indicating that the GPU memory subsystem and clock infrastructure remained active. The divergence between reported SM utilization and pinned clock state is characteristic of cryptomining code paths that drive GPU power and memory interfaces while evading software-visible SM occupancy counters. Temperature rose only 5.4$^{\circ}$C above idle, reaching a stable 35.2$^{\circ}$C.

This combination — elevated power and maximum clock frequencies at reported zero SM utilization — constitutes a distinct anomalous signature uncharacteristic of any legitimate workload class.

\textit{DDoS:} This mode contains no compute kernel. It allocates a 1 GB GPU buffer and a 1 GB pinned CPU buffer, then loops a single CUDA DMA transfer. Pinned (page-locked) memory is the key detail: it lets the PCIe DMA engine move data directly between VRAM and RAM without CPU or SM involvement. Because no CUDA kernels execute, SM utilization is 0\% and the SM clock parks at its minimum (210 MHz). Power drops to 33.6 W (below the idle baseline of 44.6 W) because with nothing to compute, only the memory controller and PCIe interface are active. The data exfiltration workload is anomalously quiet in power and temperature. Detection requires combining sub-idle power with elevated NVML-reported PCIe TX (603 KB/s mean per NVML telemetry). Neither signal alone is sufficient; the combination is uniquely diagnostic.

\textit{RowHammer:} The resulting telemetry signature is unique among all six classes: 100\% SM utilization (unlike miner, inference, or DDoS) but only 237 W (versus training's 349 W at equivalent SM saturation) because the workload is constrained by memory bandwidth rather than compute throughput. The GPU temperature ramps from 39$^{\circ}$C to 65$^{\circ}$C over 300 seconds (CoV 11.46\%, highest of all classes), consistent with progressive thermal accumulation from sustained high-frequency memory access — noting that NVML temperature reflects the GPU die sensor rather than DRAM temperature directly.

\subsection{Anomaly Detection Using Hardware Counters and Sensors}

Hardware activity counters and thermal sensors provide an efficient way to track the underlying activity in the edge AI layers \cite{r27,r28}.
Figure 2 plots all experimented workloads in two feature projections. Left panel (SM utilization vs. power): three clusters are immediately visible — Training (100\% SM util, 349 W), RowHammer (100\% SM util, 237 W), and a dense lower-left bundle containing the remaining four classes (Idle, Inference, Miner, DDoS) all at 0\% reported SM utilization. Right panel (NVML PCIe TX vs. power) resolves the lower-left bundle: Miner is unique at 41 KB/s PCIe TX; Inference (627 KB/s), DDoS (603 KB/s), and RowHammer (752 KB/s) all show elevated PCIe TX, separating from Miner by over an order of magnitude. Combined across both panels, all six classes are discriminable.

\begin{figure}[!t]
\centering
\includegraphics[width=\columnwidth]{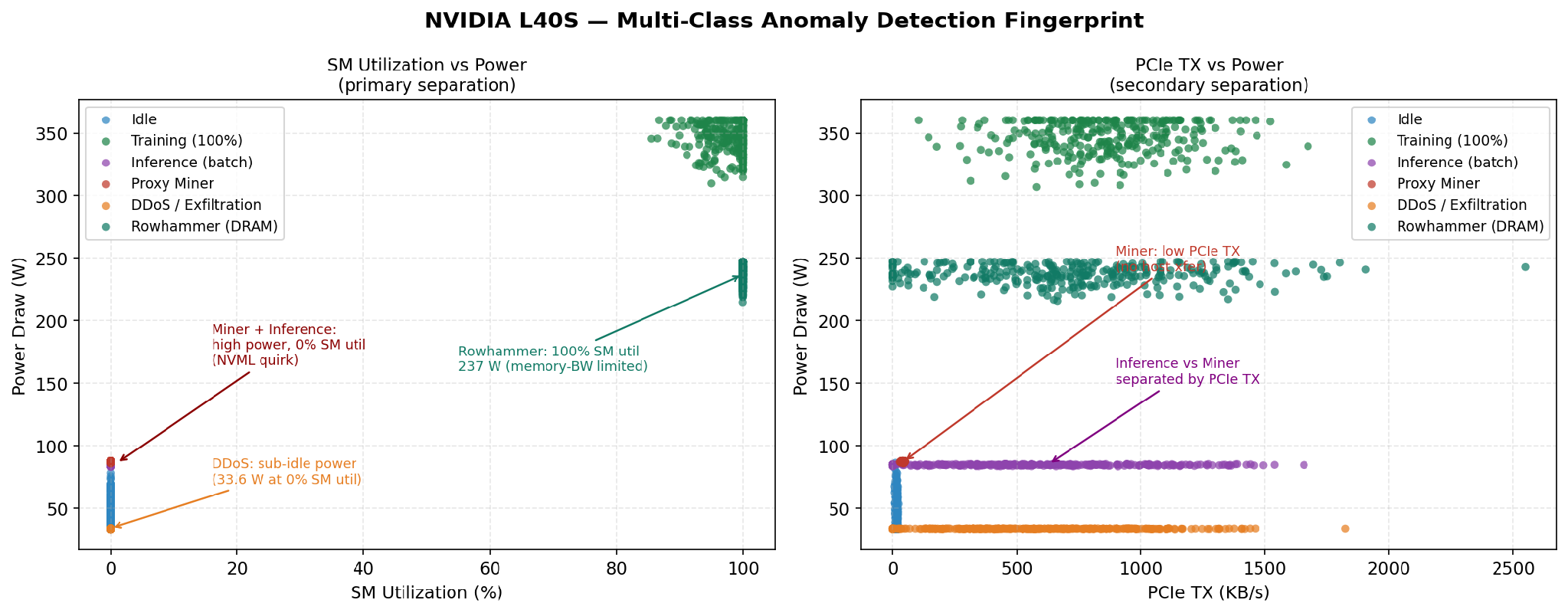}
\caption{Anomaly fingerprint. Left: SM utilization vs. power. Right: NVML PCIe TX vs. power. Six-class separation.}
\label{fig2}
\end{figure}

Figure 3 summarizes the key metrics across all workload classes (mean $\pm$ std). The classes occupy largely non-overlapping regions in this metric space. The SM Clock panel is particularly diagnostic: DDoS parks at 210 MHz (fully parked, DMA-only), while all compute-active workloads operate at 1478--2520 MHz.

\begin{figure}[!t]
\centering
\includegraphics[width=\columnwidth]{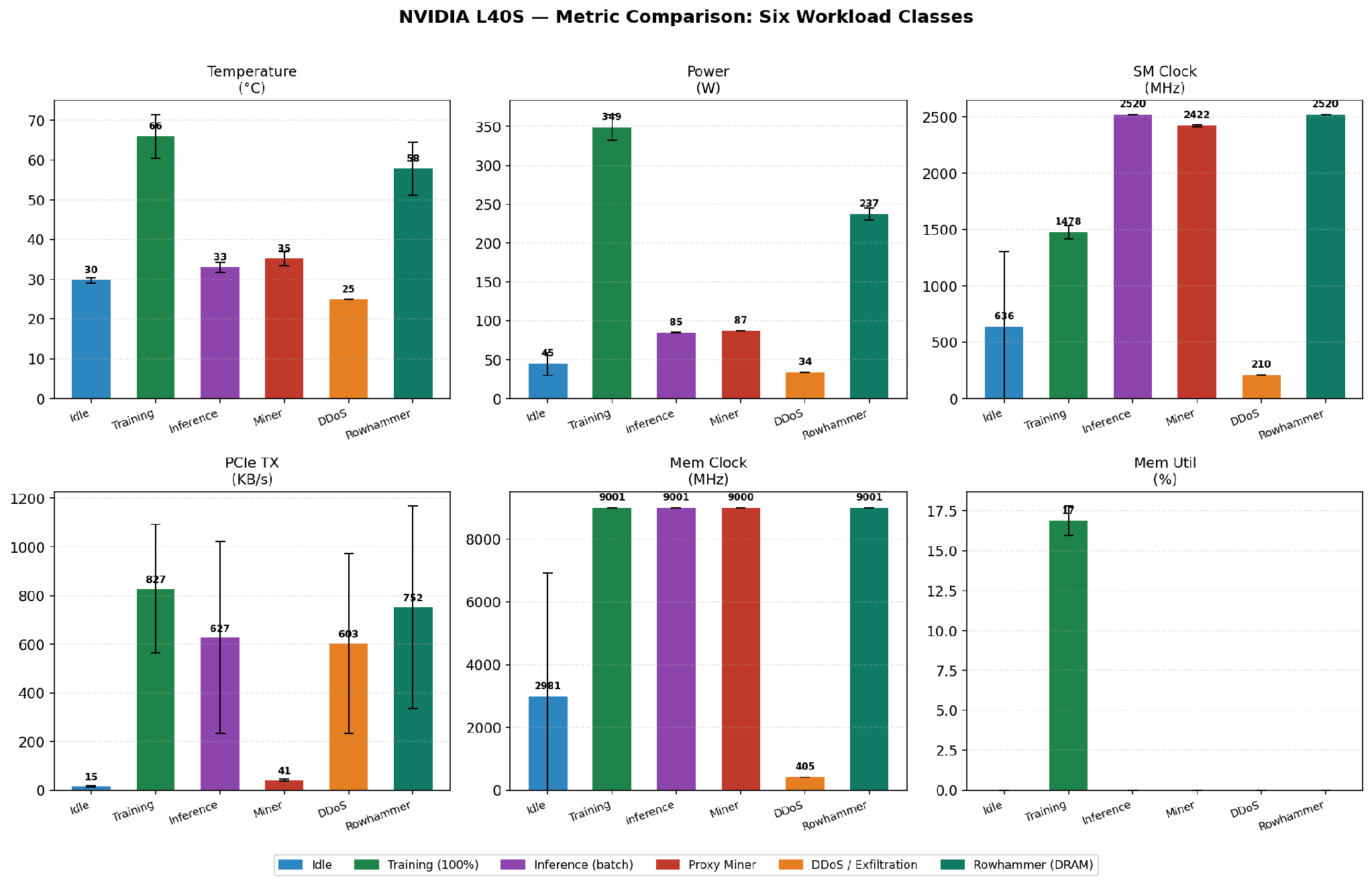}
\caption{Metric comparison of the normal and anomalous workloads.}
\label{fig3}
\end{figure}

Figure 4 shows the 300-second temperature and power traces for all six workload classes. Training ramps from 32$^{\circ}$C to 70$^{\circ}$C then stabilizes near 66$^{\circ}$C. RowHammer (teal) shows a monotonic thermal ramp from 39$^{\circ}$C to 65$^{\circ}$C throughout the full 300 s (no plateau), consistent with progressive thermal accumulation from sustained memory-intensive operation. DDoS (orange) sits flat at 25$^{\circ}$C, below the idle baseline of 29.8$^{\circ}$C. In the power panel, three distinct tiers are immediately visible: Training at 349 W, RowHammer at 237 W, and the four remaining workloads clustered below 90 W.

\begin{figure}[!t]
\centering
\includegraphics[width=\columnwidth]{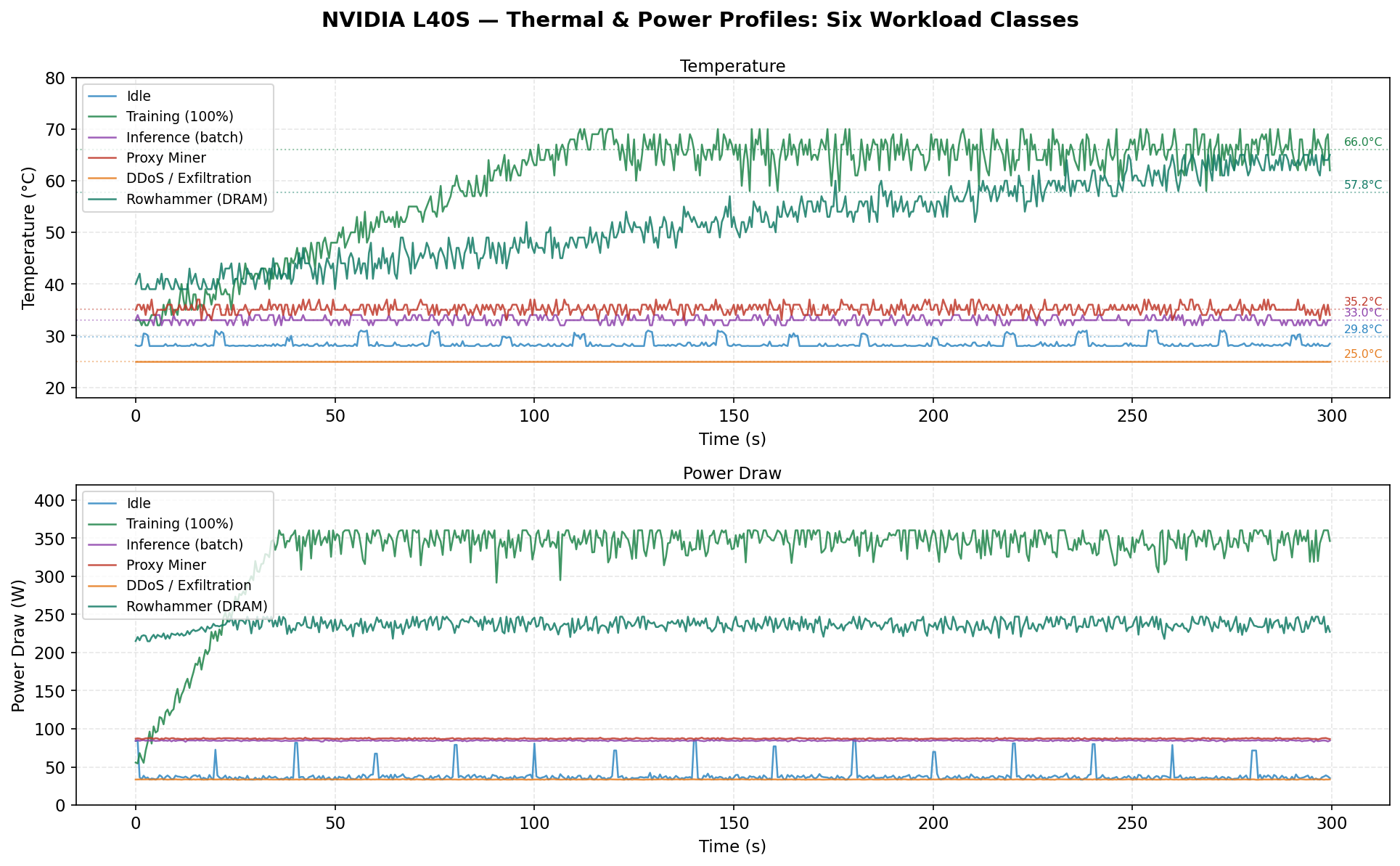}
\caption{Temperature (top) and power (bottom) time series for all six workload classes on NVIDIA L40S over 300 s.}
\label{fig4}
\end{figure}

Figure 5 shows the layered feature hierarchy across four projections. Layer 4 (Temperature vs. SM Clock, bottom-right) achieves near-complete separation of all six classes when using mean or steady-state values: DDoS isolated at 25$^{\circ}$C / 210 MHz; Idle scattered across low-clock P-states; Inference and RowHammer lock at 2520 MHz while Miner clusters near 2422 MHz (range 2370--2520 MHz), all at distinct temperatures; Training at 1478 MHz mean / 50--70$^{\circ}$C. It is important to note that training begins cold at 32$^{\circ}$C with an initial SM clock near 2520 MHz, transiently overlapping the Inference region during the first $\sim$30 s of ramp-up. Single-projection separation is therefore not perfect for raw per-sample point clouds; aggregate features or a classifier that handles startup transients achieves full separation.

\begin{figure}[!t]
\centering
\includegraphics[width=\columnwidth]{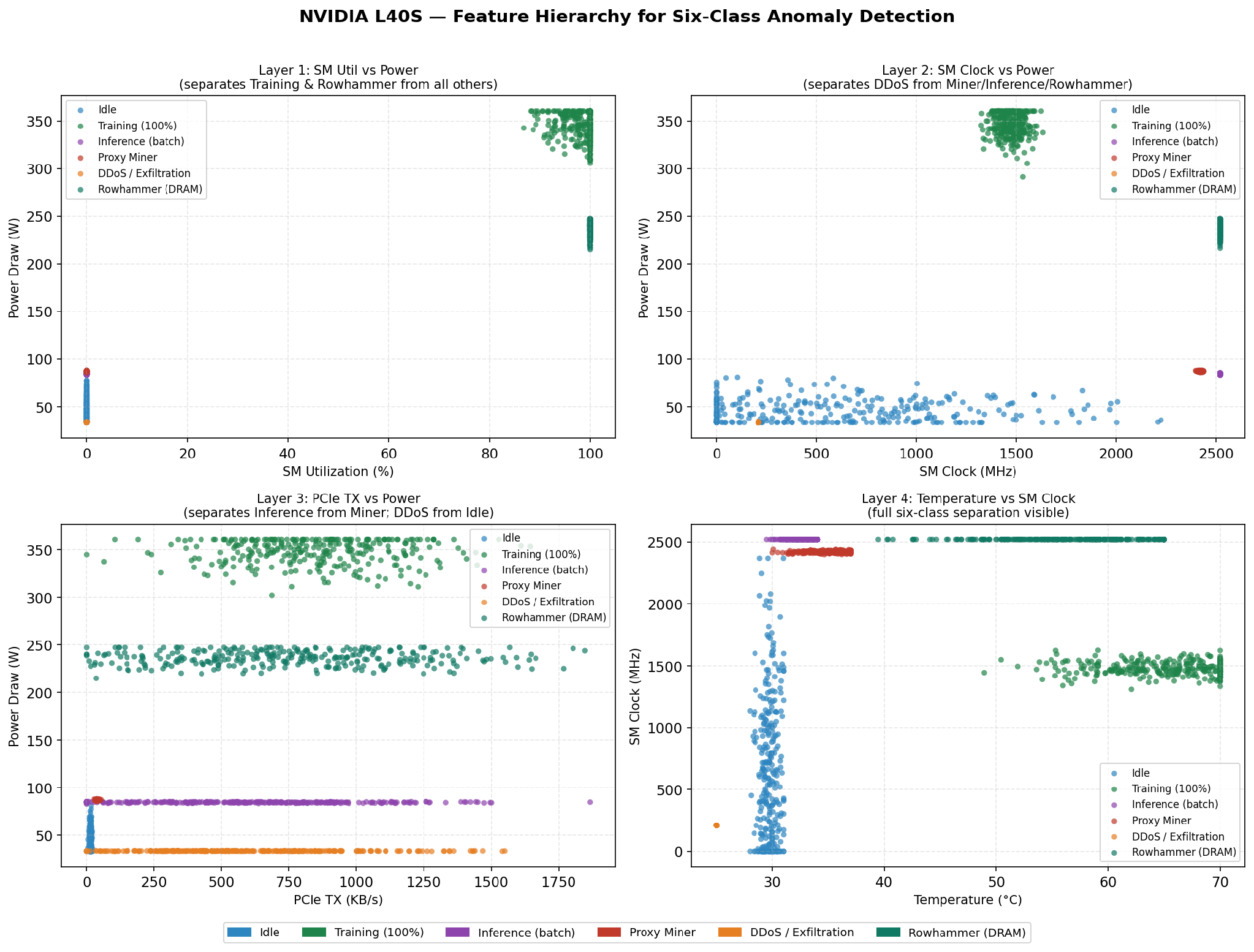}
\caption{Feature hierarchy: four projections showing layered discrimination from SM util $\rightarrow$ power $\rightarrow$ PCIe TX $\rightarrow$ temperature vs.\ SM clock.}
\label{fig5}
\end{figure}

While baseline hardware anomaly detection, using performance counters and physical sensors, may exhibit coverage gaps in detection for stealthy, low-footprint threats like passive data exfiltration or silent backdoors, the broader architecture of the proposed stacked hardware security controller \cite{r19} integrates semantic activity tracking and supplementary verification mechanisms designed to capture such subtle abnormalities. Although a comprehensive empirical evaluation of these advanced semantic modules falls outside the scope of the current study, their integration underscores the extensible nature of the proposed framework beyond conventional microarchitectural monitoring.

\section{Conclusions}

Edge AI systems face a growing number and diversity of cyberattacks in the field. Software-based malware detection in edge AI systems is considerably more challenging because of limited energy resources, battery capacity, thermal budgets, computational capabilities, and communication bandwidths. As a result, conventional software-based cybersecurity solutions are often impractical or cause serious issues during deployments.

In this paper, we propose a stacked cybersecurity controller unit that continuously monitors and analyzes the underlying activity from hardware counters and on-chip sensors. We show that this low-overhead hardware monitoring approach effectively discriminates between normal activity (including high-compute training periods) and anomalous activity, with distinct cluster separation across six workload classes that enables accurate anomaly classification.


\end{document}